\documentclass[12pt,a4paper]{article}
\usepackage[margin=25mm]{geometry}
\usepackage{amsmath,amssymb,mathtools,graphicx,booktabs,microtype,hyperref}

\hypersetup{hidelinks}\graphicspath{{figures/}}

\newcommand{\dd}{\mathrm d}\newcommand{\Ree}{\operatorname{Re}}
\title{Theoretical Description of Electronic Modulation and Superconducting Coherence}
\author{Jaehwahn Kim$^{1}$, Davis A. Rens$^{2,3}$,\\
Waqas Khalid$^{2,3}$, and Hyunchul Kim$^{1}$\\[0.6em]
\normalsize $^{1}$Hyunsung T\&C Laboratory, Suwon 16679, Republic of Korea\\
\normalsize $^{2}$Department of Physics, University of California, Berkeley, California 94720, USA\\
\normalsize $^{3}$SuperMatics, Inc., Berkeley, California 94704, USA}
\date{October 19, 2026}
\begin{document}\maketitle
\begin{abstract}
An electronic modulation can retain a substantial local amplitude while its translational coherence is limited by spatial disorder. Its coupling to superconductivity therefore cannot be characterized by a single modulation intensity. We develop a phenomenological formulation of a finite-wave-vector complex electronic envelope coupled to a charged superconducting field. Translation and gauge symmetries constrain the interaction terms, while amplitude, modulation-gradient and condensate-gradient couplings act on distinct responses. The resulting variational equations give uniform coexistence conditions, local stability criteria and phase-gradient coefficients. We define the helicity modulus by a boundary twist, distinguish phase-only from amplitude-weighted correlations, and give an explicit minimal spectral Hamiltonian illustrating how the electronic texture may enter momentum-resolved measurements. The proposed tests use distinct spatial, spectral and stiffness observables; none assumes that a tunnelling gap directly measures condensate amplitude. An explicit kernel expansion yields the leading displacement of the selected wave vector, while the thermodynamic twist response determines phase rigidity independently of charge intensity. The analytical relations and selected numerical responses distinguish amplitude, spatial organization and condensate rigidity within the specified coupled-field model.
\end{abstract}

\section{Introduction}
The characteristic wave vector of an electronic modulation does not uniquely describe its spatial organization. A state can possess a locally resolved periodic pattern while the modulation phase varies among finite domains. The integrated scattering intensity, peak width, local amplitude and phase correlation then describe different aspects of the same texture. Distinguishing these observables is especially important in a superconductor, whose condensate carries independent amplitude and phase degrees of freedom.

The phenomenology of high-temperature superconductivity and its pseudogap regime has been reviewed extensively \cite{timusk1999,keimer2015}. Charge modulations and superconductivity have been examined in cuprates by diffraction and scanning spectroscopy \cite{ghiringhelli2012,chang2012,fujita2014,vershinin2004}. Such observations motivate a coupled-field description, but do not determine the signs or magnitudes of the coupling constants. Neither a finite-$q$ feature nor a local gap--modulation correlation is by itself a measurement of superfluid stiffness. A phenomenology that keeps these observables distinct is needed before quantitative comparisons are attempted.

We use a finite-wave-vector envelope, the modulated electron lattice (MEL) field of Ref.~\cite{mel2025}, to represent electronic organization without imposing long-range translational order. The aim is to establish the mathematical relation between this envelope and superconducting coherence. A Ginzburg--Landau (GL) functional is used as a convenient variational language; it is not identified with the underlying microscopic origin of the modulation or of pairing.

Intertwined-order and pair-density-wave descriptions provide broader context \cite{fradkin2015,agterberg2020}. Prior coupled-order theories already include competing orders, disorder and gradient interactions \cite{hayward2014,caplan2015,yukivelson2019,morozovska2023}. The present formulation resolves the amplitude, electronic-gradient and condensate-gradient response channels within a common variational description, and extends the coexistence analysis of biquadratically coupled order parameters \cite{liufisher1973,knf1976} by the two gradient couplings and the twist response.

The distinction between a local modulation and its long-range average is relevant to diffraction and tunnelling alike. Intra-unit-cell $s$, $s'$ and $d$ form factors distinguish different oxygen and copper components, with $s'$ denoting an oxygen-bond channel rather than a copper-site order parameter \cite{fujita2014,comin2016}. Local commensurability and an incommensurate average can coexist when phase discommensurations separate short domains \cite{mesaros2016}. The carrier wave vector therefore does not exhaust the information contained in the complex envelope.

Superconductivity introduces a separate gauge degree of freedom. A translational-phase correlation length, a charge structure-factor width, and a superconducting helicity modulus are distinct mathematical observables. Their variations need not have the same sign. Earlier coupled-order theories have treated competing charge order, random pinning and field-dependent textures \cite{hayward2014,caplan2015,nie2015,kivelson2002,yukivelson2019}; the present analysis retains those precedents while identifying the response channels of the particular truncated functional below. Related mixed-gradient invariants have been considered in Landau descriptions of modulated electronic and structural order \cite{morozovska2023}.

Scanning tunnelling microscopy and spectroscopy (STM/STS) resolve a matrix-element-weighted local response \cite{fischer2007,hoffman2002,vershinin2004,wise2008,slezak2008}, whereas angle-resolved photoemission spectroscopy (ARPES) probes occupied momentum-resolved spectral weight \cite{damascelli2003,norman1998}. The field-theoretic response functions and the spectroscopic measurement operators are specified separately.

\section{Electronic Modulation and Symmetry}
\subsection{Finite-Momentum Envelope}
A real density perturbation can be expressed as
\begin{equation}\label{density}\delta\rho(\mathbf r)=\Ree\{\Phi(\mathbf r)f(\mathbf r)e^{i\mathbf Q\cdot\mathbf r}\}+\delta\rho_{\mathrm{harm}}(\mathbf r),\quad \Phi=Ae^{i\varphi}.
\end{equation}
Here $f$ is the intra-unit-cell form factor, $A\ge0$ is the slowly varying envelope amplitude, $\varphi$ is its translational phase and $\delta\rho_{\mathrm{harm}}$ collects higher harmonics. Where the envelope is sufficiently smooth and nonzero, a local effective wave vector can be defined by $\mathbf q_{\mathrm{loc}}=\mathbf Q+\nabla\varphi$.

Translating the pattern by $\mathbf a$ maps $\Phi(\mathbf r)\to e^{-i\mathbf Q\cdot\mathbf a}\Phi(\mathbf r-\mathbf a)$; the inverse convention flips the sign, and no invariant below depends on it. For generic incommensurate $\mathbf Q$, a homogeneous term proportional to $\Phi|\psi|^2$ is not translation invariant. The lowest local amplitude coupling is instead $|\Phi|^2|\psi|^2$. At commensurability, when $n\mathbf Q$ is reciprocal, Umklapp terms $\Phi^n+\Phi^{*n}$ can appear; we exclude them from the generic model.

The real modulation field used in the earlier MEL phenomenology is related to the present envelope through the physical density component $\delta\rho_{\mathbf Q}(\mathbf r)=\operatorname{Re}[\Phi(\mathbf r)f(\mathbf r)e^{i\mathbf Q\cdot\mathbf r}]$. A local coupling linear in this oscillatory density, $\delta\rho_{\mathbf Q}|\psi|^2$, and the translationally invariant envelope coupling $|\Phi|^2|\psi|^2$ are different operators: coarse graining over the carrier suppresses the former in a homogeneous incommensurate background, while the latter survives. The first version of Ref.~\cite{mel2025} used the former coupling; for incommensurate $\mathbf Q$ it is not translation invariant, and the present functional replaces it by $|\Phi|^2|\psi|^2$.

The superconducting order parameter is $\psi=Se^{i\theta}$, with $S\ge0$. Its gauge-covariant derivative is
\begin{equation}\label{D}D_j\psi=(\partial_j-iq_s A_j/\hbar)\psi,\qquad q_s=2e.\end{equation}
Here $e<0$ is the signed electron charge. The scalar amplitudes $A,S$ are distinguished from the vector potential $\mathbf A$ and structure factor $S_\Phi(\mathbf k)$; the scalar twist $\delta$, carrier displacement $\boldsymbol\delta$ and variational prefixes $\delta$ are distinguished by their arguments.
The envelope phase $\varphi$ and condensate phase $\theta$ obey distinct translation and gauge transformation laws. In a tetragonal plane two orientations $\Phi_x,\Phi_y$ may be needed. The single-envelope sector describes one selected modulation orientation; symmetry-related orientations can be incorporated as additional complex fields.

\begin{table}[!htbp]\centering\small
\caption{Notation for the modulated electronic envelope and superconducting field.}\label{tab:field_symbols}
\begin{tabular}{@{}p{.23\linewidth}p{.67\linewidth}@{}}\toprule
Symbol & Definition and convention\\\midrule
$\mathbf r,\mathbf Q$ & Position and chosen carrier for the slowly varying envelope; the carrier is a representation convention, not necessarily the physical minimum.\\
$\delta\rho,\delta\rho_{\rm harm}$ & Real charge-density modulation and contributions from higher harmonics.\\
$\Phi=Ae^{i\varphi}$ & Complex slowly varying envelope; $A\geq0$ is its amplitude and $\varphi$ its translational phase.\\
$f(\mathbf r)$ & Intra-unit-cell form factor fixed by the chosen charge channel.\\
$\mathbf q_{\rm loc}$ & Local wave vector $\mathbf Q+\nabla\varphi$ where the envelope is smooth and nonzero.\\
$\psi=Se^{i\theta}$ & Superconducting order parameter with amplitude $S$ and condensate phase $\theta$.\\
$D_j,\mathbf A,q_s$ & Gauge-covariant derivative, electromagnetic vector potential and pair charge $q_s=2e$.\\
\bottomrule\end{tabular}\end{table}

The two phases in Table~\ref{tab:field_symbols} are not interchangeable. The electronic phase describes translations of a density modulation, whereas the superconducting phase transforms under electromagnetic gauge transformations. A term that couples their gradients must satisfy both symmetries; such a term cannot be inferred merely from the fact that both fields are complex. This distinction will also matter when comparing spatial demodulation with an independently measured superfluid response.

\subsection{Symmetry-Allowed Interactions}
The retained couplings are
\begin{align}\label{invariants}
I_0&=|\Phi|^2|\psi|^2,\\
I_1&=|\psi|^2\sum_j|\partial_j\Phi|^2,\\
I_2&=|\Phi|^2\sum_j|D_j\psi|^2.
\end{align}
A mixed amplitude-gradient invariant such as $\nabla|\Phi|^2\cdot\nabla|\psi|^2$ is also symmetry allowed. Omitting it is a truncation, not a symmetry prohibition. One first-order invariant is also allowed, $g_1|\psi|^2\,\hat{\mathbf Q}\cdot\mathrm{Im}(\Phi^*\nabla\Phi)$; it is even under time reversal and inversion and is the real-space form of the $g_1$ term in Eq.~\eqref{qexample}. It shifts the selected wave vector along $\hat{\mathbf Q}$ by $-S^2g_1/[2(K_m+\kappa S^2)]$. No symmetry forces $g_1=0$ at generic incommensurate $\mathbf Q$; its omission in Eq.~\eqref{F} is a truncation.

A current--current invariant $\mathbf J^\Phi\!\cdot\mathbf J^\psi$, with $\mathbf J^\Phi=\mathrm{Im}(\Phi^*\nabla\Phi)$ and $\mathbf J^\psi=\mathrm{Im}(\psi^*\mathbf D\psi)$, is odd under time reversal: the static density is even, so $\Phi\to\Phi$, while $\psi\to\psi^*$ and $\mathbf A\to-\mathbf A$. It is also odd under inversion and is excluded in zero field.

\section{Variational Formulation}
\subsection{Free-Energy Functional}
On a two-dimensional domain $\Omega$ we define
\begin{align}\label{F}
F=\int_\Omega\dd^2r\,\big\{&a_s|\psi|^2+\tfrac{b_s}{2}|\psi|^4+K_s|\mathbf D\psi|^2\nonumber\\
&+a_m|\Phi|^2+\tfrac{b_m}{2}|\Phi|^4+K_m|\nabla\Phi|^2\nonumber\\
&+\gamma|\Phi|^2|\psi|^2+\kappa|\psi|^2|\nabla\Phi|^2\nonumber\\
&+\kappa_\psi|\Phi|^2|\mathbf D\psi|^2-\Ree[h^*(\mathbf r)\Phi]\big\}.
\end{align}
The parameters $a_s,a_m$ can depend on temperature; $b_s,b_m>0$ are local quartic coefficients. The optional field $h$ represents quenched pinning.

The reference carrier $\mathbf Q_0$ is chosen at a stationary minimum of the quadratic full-density kernel in the normal reference state. In a fixed-carrier expansion the physical wave vector is $\mathbf q=\mathbf Q_0+\mathbf k$. Writing $\alpha_0(\mathbf q)$ for the reference quadratic kernel, its expansion is
\begin{equation}\label{qkernel}
\alpha_0(\mathbf Q_0+\mathbf k)=a_m+\tfrac12 k_i H_{ij}k_j+O(|\mathbf k|^3),\quad H_{ij}=\left.\partial_{q_i}\partial_{q_j}\alpha_0\right|_{\mathbf Q_0}.
\end{equation}
The first derivative vanishes because $\mathbf Q_0$ is stationary. For an isotropic curvature $H_{ij}=2K_m\delta_{ij}$, the quadratic contribution is $K_m|\nabla\Phi|^2$. The factor of two is fixed by this convention. In an anisotropic material the correct term is $\tfrac12 H_{ij}(\partial_i\Phi)^*(\partial_j\Phi)$, not a scalar $K_m$ unless isotropy is assumed. The envelope approximation requires $|\mathbf k|$ to remain within the domain where the Taylor expansion is controlled. A change of carrier convention must transform the gradient operator as shown below; it is not a physical change of the ordering wave vector.

\begin{table}[!htbp]\centering\small
\caption{Wave-vector conventions and kernel derivatives.}\label{tab:q_symbols}
\begin{tabular}{@{}p{.25\linewidth}p{.65\linewidth}@{}}\toprule
Symbol & Definition\\\midrule
$\mathbf Q_0$ & Reference carrier chosen at a stationary minimum of $\alpha_0$; fixed when taking temperature derivatives.\\
$\mathbf q_*(T)$ & Temperature-dependent minimizer of the effective quadratic kernel, when it exists uniquely.\\
$\mathbf k$ & Deviation $\mathbf q-\mathbf Q_0$ from the reference carrier.\\
$\mathbf q_{\rm loc}$ & Local phase-gradient wave vector $\mathbf Q+\nabla\varphi$ away from amplitude zeros.\\
$\alpha_0(\mathbf q)$ & Quadratic electronic kernel in the chosen reference state.\\
$H_{ij}$ & Hessian of $\alpha_0$ at $\mathbf Q_0$; positive definite for an isolated stable minimum.\\
$K_m$ & Isotropic envelope-gradient coefficient $H_{ii}/2$ (no sum), in the stated normalization.\\
\bottomrule\end{tabular}\end{table}

\begin{table}[!htbp]\centering\small
\caption{Coefficients and terms of the truncated free-energy functional in Eq.~\eqref{F}.}\label{tab:energy_symbols}
\begin{tabular}{@{}p{.23\linewidth}p{.67\linewidth}@{}}\toprule
Symbol & Definition and convention\\\midrule
$F,\Omega$ & Free energy and two-dimensional integration domain.\\
$a_s,a_m$ & Quadratic coefficients in the superconducting and electronic sectors.\\
$b_s,b_m$ & Positive quartic self-interaction coefficients.\\
$K_s,K_m$ & Bare condensate and electronic-envelope gradient coefficients.\\
$\gamma$ & Local biquadratic amplitude coupling between the two sectors.\\
$\kappa$ & Electronic-envelope gradient coupling weighted by condensate amplitude.\\
$\kappa_\psi$ & Condensate-gradient coupling weighted by electronic-envelope amplitude.\\
$h(\mathbf r)$ & Optional complex pinning field acting linearly on the envelope.\\
\bottomrule\end{tabular}\end{table}

Equation~\eqref{F} is a low-order, symmetry-compatible truncation. Its coefficients are phenomenological: a positive or negative value of a coupling cannot be assigned to a specific material without a separate microscopic or experimental determination. In particular, the gradient couplings affect the spatial cost of textures, not merely the uniform amplitudes. The local pinning field is included so that disorder can be represented mathematically; the statistical distribution of the pinning field enters the corresponding disorder-averaged correlation functions.

\subsection{Temperature Dependence of the Selected Wave Vector}
A temperature-dependent modulation wave vector is permitted by the same variational framework, but it is not forced to change. To establish the precise condition, introduce the effective quadratic kernel for a homogeneous superconducting background $S^2(T)=|\psi(T)|^2$:
\begin{equation}\label{alphaT}
\alpha_{\rm eff}(\mathbf q,T)=\alpha_0(\mathbf q,T)+S^2(T)g(\mathbf q,T).
\end{equation}
Here $g$ is a symmetry-allowed, potentially momentum-dependent quadratic coupling. The constant part of $g$ includes the uniform amplitude interaction $\gamma$, whereas the curvature of $g$ encodes wave-vector-sensitive coupling. The minimal truncated functional in Eq.~\eqref{F}, with the carrier fixed at $\mathbf Q_0$, has a constant $\gamma$ and an isotropic quadratic gradient correction $\kappa S^2|\mathbf q-\mathbf Q_0|^2$. In that restricted model, and with a temperature-independent $\alpha_0$, the preferred carrier remains at $\mathbf Q_0$ as long as the total curvature remains positive. A shift from the first-order invariant of Sec.~2.2 is omitted by this truncation, not excluded by symmetry.

For a general smooth kernel the preferred wave vector satisfies
\begin{equation}\label{qstationary}
\nabla_{\!q}\alpha_{\rm eff}(\mathbf q_*(T),T)=0,\qquad \mathcal H_{ij}=\partial_{q_i}\partial_{q_j}\alpha_{\rm eff}|_{\mathbf q_*}.
\end{equation}
If the Hessian $\mathcal H$ is invertible, differentiating this stationarity condition with respect to temperature gives the exact local identity
\begin{equation}\label{qderivative}
\frac{d q_{*,i}}{dT}=-(\mathcal H^{-1})_{ij}\left.\partial_T\partial_{q_j}\alpha_{\rm eff}(\mathbf q,T)\right|_{\mathbf q=\mathbf q_*(T)}.
\end{equation}
At leading order in a weak coupling $S^2g$ about a temperature-independent $\mathbf Q_0$,
\begin{equation}\label{qshift}
\delta\mathbf q_*(T)=-S^2(T)H^{-1}\nabla_{\!q}g(\mathbf Q_0,T)+O(S^4),
\end{equation}
where the $O(S^4)$ remainder presumes a smooth perturbative expansion and a temperature-independent reference kernel $\alpha_0$. If $\alpha_0$ depends on temperature, its shift of the minimum must be calculated separately. A nonzero shift requires a nonzero momentum derivative of the coupling or an intrinsic movement of the bare minimum. Its sign and magnitude are material dependent. At a degenerate or non-smooth minimum, Eq.~\eqref{qderivative} does not apply and the minimizing wave vector may change discontinuously.

For an explicit one-dimensional local kernel, let $k=q-Q_0$ and consider
\begin{equation}\label{qexample}
\alpha_{\rm eff}(k,T)=a(T)+\frac{H}{2}k^2+S^2(T)[\gamma+g_1 k+\tfrac12 g_2k^2],
\end{equation}
with $H+S^2g_2>0$. Its unique quadratic minimum is
\begin{equation}\label{qexample_result}
q_*(T)-Q_0=-\frac{S^2(T)g_1}{H+S^2(T)g_2},\qquad
\frac{\dd q_*}{\dd T}=-\frac{H g_1}{[H+S^2(T)g_2]^2}\frac{\dd S^2}{\dd T},
\end{equation}
for temperature-independent $H$, $g_1$ and $g_2$. The $g_1=0$ limit retains the original carrier, whereas a nonzero $g_1$ shifts it continuously while the curvature stays positive. This explicit solution supplements the general implicit-function relation in Eq.~\eqref{qderivative}. The $g_1$ term represents a momentum-asymmetric correction around the selected orientation; it is symmetry allowed at generic incommensurate $\mathbf Q$, although additional symmetry can require $g_1=0$ in a particular model.

\begin{table}[!htbp]\centering\small
\caption{Quantities in the temperature-dependent wave-vector calculation.}\label{tab:qtemp_symbols}
\begin{tabular}{@{}p{.27\linewidth}p{.63\linewidth}@{}}\toprule
Symbol & Definition\\\midrule
$S^2(T)$ & Homogeneous condensate magnitude squared, a model-dependent function of temperature.\\
$g(\mathbf q,T)$ & Momentum-dependent quadratic coupling to superconductivity; not fixed by symmetry alone.\\
$\alpha_{\rm eff}$ & Total quadratic kernel whose stable minimum determines $\mathbf q_*(T)$.\\
$\mathcal H$ & Hessian of $\alpha_{\rm eff}$ at its temperature-dependent minimum.\\
$\delta\mathbf q_*$ & Shift of the minimizing wave vector from $\mathbf Q_0$ in the perturbative regime.\\
\bottomrule\end{tabular}\end{table}

Equation~\eqref{qderivative} provides a testable theoretical relation between a measured wave-vector trajectory and a specified microscopic or phenomenological kernel. The trajectory is fixed by the temperature and momentum derivatives of $\alpha_0$, $g$ and $S(T)$ in the chosen effective kernel. The measured local distribution $P(\mathbf q_{\rm loc}|T)$ need not, however, peak at the homogeneous kernel minimum when pinning, finite domains or discommensurations are present. The temperature series and spatially resolved observables therefore test complementary aspects of the present formulation.

\subsection{Coupled Euler--Lagrange Equations}
Varying the fields and their complex conjugates independently, with boundary terms set to zero, gives
\begin{align}\label{ELs}
0={}&(a_s+b_s|\psi|^2+\gamma|\Phi|^2+\kappa|\nabla\Phi|^2)\psi\nonumber\\
&-\mathbf D\cdot[(K_s+\kappa_\psi|\Phi|^2)\mathbf D\psi],
\end{align}
\begin{align}\label{ELm}
0={}&(a_m+b_m|\Phi|^2+\gamma|\psi|^2+\kappa_\psi|\mathbf D\psi|^2)\Phi\nonumber\\
&-\nabla\cdot[(K_m+\kappa|\psi|^2)\nabla\Phi]-\tfrac12 h.
\end{align}
The equations are coupled even when one field has no imposed spatial modulation: a gradient of either field contributes to the effective local mass of the other.

The polar decompositions give the exact identities
\begin{align}\label{polar}
|\nabla\Phi|^2&=|\nabla A|^2+A^2|\nabla\varphi|^2,\\
|\mathbf D\psi|^2&=|\nabla S|^2+S^2|\nabla\theta-q_s\mathbf A/\hbar|^2.
\end{align}
Thus the $\kappa$ term affects both amplitude gradients and phase gradients of the electronic envelope. Calling it a purely phase coupling is justified only after a separate approximation in which amplitude gradients are negligible.
\section{Homogeneous Coexistence and Stability}
With $h=\mathbf A=0$ and uniform fields, let $x=S^2$ and $y=A^2$. The local energy is
\begin{equation}\label{uniform}f_0=a_sx+a_my+\tfrac12 b_sx^2+\tfrac12 b_my^2+\gamma xy.\end{equation}
An interior stationary point obeys
\begin{equation}\label{matrix}\begin{pmatrix}b_s&\gamma\\\gamma&b_m\end{pmatrix}\binom{x}{y}=-\binom{a_s}{a_m}.\end{equation}
For $D=b_sb_m-\gamma^2\ne0$, the solution is
\begin{equation}\label{solution}x=\frac{\gamma a_m-b_m a_s}{D},\qquad y=\frac{\gamma a_s-b_s a_m}{D}.\end{equation}
Both amplitudes must be positive to describe an interior coexistence state. Let $\mathbf u=(x,y)^T$, $\mathbf u_*=(x_*,y_*)^T$ and $B=\begin{psmallmatrix}b_s&\gamma\\\gamma&b_m\end{psmallmatrix}$. For $b_s>0$ and $D>0$, $B$ is positive definite. Since the local potential is quadratic in $\mathbf u$, stationarity implies the exact identity
\begin{equation}\label{globalmin}
f_0(\mathbf u)-f_0(\mathbf u_*)=\tfrac12(\mathbf u-\mathbf u_*)^TB(\mathbf u-\mathbf u_*)>0\quad(\mathbf u\ne\mathbf u_*).
\end{equation}
Consequently, when $x_*,y_*>0$, the coexistence solution is the unique global minimizer of the uniform local potential on the physical quadrant. This result does not establish stability of spatially varying configurations, which is governed additionally by the gradient sector.

For $b_s,b_m>0$ and $\gamma> -\sqrt{b_sb_m}$, the quartic form is strictly positive away from the origin in the physical quadrant \cite{liufisher1973,knf1976}. At $\gamma=-\sqrt{b_sb_m}$ the quartic term vanishes along one nonnegative ray; boundedness of the full local potential then depends on the quadratic coefficients along that ray. Large positive $\gamma$ does not cause quartic unboundedness, although it can preclude an interior coexistence minimum.

\begin{table}[!htbp]\centering\small
\caption{Uniform coexistence and stability notation for Eqs.~\eqref{uniform}--\eqref{solution}.}\label{tab:uniform_symbols}
\begin{tabular}{@{}p{.23\linewidth}p{.67\linewidth}@{}}\toprule
Symbol & Definition and convention\\\midrule
$x=S^2,\ y=A^2$ & Nonnegative squared magnitudes of the two uniform fields.\\
$f_0$ & Local uniform free-energy density with gradients and pinning removed.\\
$D=b_sb_m-\gamma^2$ & Determinant of the uniform amplitude-response matrix.\\
$b_s,b_m,\gamma$ & Quartic self-interactions and cross-sector amplitude coupling.\\
$K_s^{\rm eff},K_m^{\rm eff}$ & Background-dependent gradient coefficients $K_s+\kappa_\psi|\Phi|^2$ and $K_m+\kappa|\psi|^2$.\\
\bottomrule\end{tabular}\end{table}

\begin{figure}[tb]\centering\includegraphics[width=.78\linewidth]{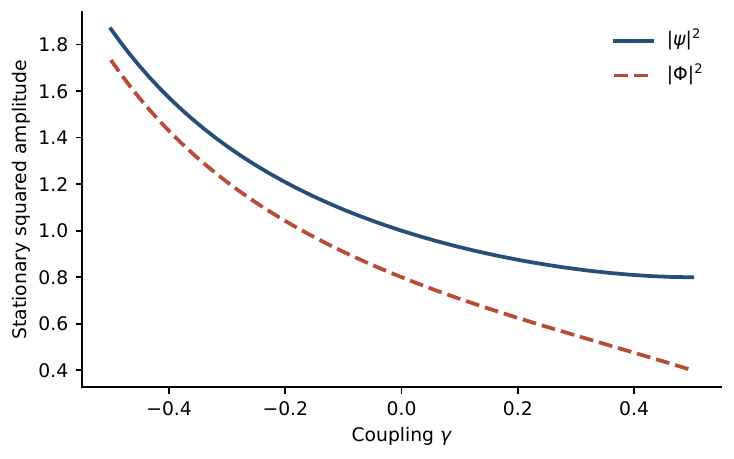}\caption{Exact uniform coexistence branch of the coupled-field model, with $a_s=-1$, $a_m=-0.8$, $b_s=b_m=1$ and variable $\gamma$. The curves are $S^2=(-a_s b_m+\gamma a_m)/(b_sb_m-\gamma^2)$ and $A^2=(-a_m b_s+\gamma a_s)/(b_sb_m-\gamma^2)$ where both amplitudes are positive. All quantities are dimensionless.}\label{fig:coexistence}\end{figure}

Figure~\ref{fig:coexistence} displays the analytic coexistence branch. Within the positive-definite domain of Eq.~\eqref{globalmin}, its physical interior is the global minimum of the uniform potential. Outside this domain, the stable uniform state is determined by the boundary solutions and the appropriate higher-order stabilization.

\subsection{Uniform-Limit Stiffness Response}
The homogeneous coexistence branch permits a closed comparison of its bare phase-gradient response with a decoupled superconducting reference. For $x=|\psi|^2>0$, $y=|\Phi|^2>0$, $x_0=-a_s/b_s>0$ and $K_s>0$, the stationary condition gives $x=x_0-\gamma y/b_s$. Hence
\begin{equation}\label{main_enhancement}
R_{\rm bare}=\frac{x}{x_0}\left(1+\frac{\kappa_\psi y}{K_s}\right),\qquad
R_{\rm bare}>1\ \Longleftrightarrow\ \kappa_\psi x>\frac{K_s\gamma}{b_s},
\end{equation}
where $R_{\rm bare}=\Upsilon_{\rm bare}/\Upsilon_{0,\rm bare}$ and both the interior coexistence and local gradient-stability conditions are imposed. For $\kappa_\psi<0$ the continuum functional is unbounded below (Appendix~\ref{app:stability}), so that part of the domain of Eq.~\eqref{main_enhancement} applies only with a lattice cutoff or an appropriate bounded-amplitude restriction. Its detailed derivation is in Appendix~\ref{app:stiffness}. The thermodynamic helicity modulus, including spatial fluctuations, is defined separately in Sec.~\ref{sec:helicity}.

With the simulation parameters of Appendix~\ref{app:lattice}, Eq.~\eqref{solution} gives $x/x_0=1.059$ and $0.964$, and $y/y_0=1.254$ and $0.769$, at $\gamma=-0.3$ and $+0.3$, respectively, where $y_0=-a_m/b_m$. At $\kappa=\kappa_\psi=0$, Eq.~\eqref{main_enhancement} then gives the same bare stiffness ratios $x/x_0$, compared with Monte Carlo stiffness ratios $1.07$ and $0.96$ and amplitude ratios $1.29$ and $0.72$ in Fig.~\ref{fig:mc}.

\subsection{Gradient Stability Conditions}
The stability of the homogeneous potential and that of the spatial functional are separate properties. In the quadratic-gradient truncation, the gradient energy takes the form
\begin{equation}\label{grad_full}
F_{\rm grad}=\int_\Omega\!\dd^2r\,\big[(K_m+\kappa|\psi|^2)|\nabla\Phi|^2+(K_s+\kappa_\psi|\Phi|^2)|\mathbf D\psi|^2\big].
\end{equation}
For a prescribed bounded-amplitude sector, the pointwise inequalities
\begin{equation}\label{Keff}
K_m+\kappa|\psi|^2\geq0,\qquad K_s+\kappa_\psi|\Phi|^2\geq0
\end{equation}
ensure nonnegative effective gradient coefficients $K_m^{\rm eff}\equiv K_m+\kappa|\psi|^2$ and $K_s^{\rm eff}\equiv K_s+\kappa_\psi|\Phi|^2$, and hence nonnegative quadratic gradient energy. At a uniform coexistence solution, strict inequalities in Eq.~\eqref{Keff} give positive leading gradient costs for infinitesimal phase and amplitude variations. The full second variation is still needed for an inhomogeneous stationary background.

For the unbounded-amplitude continuum functional of Eq.~\eqref{F}, with $K_s,K_m>0$ and strictly coercive quartic potential $\gamma> -\sqrt{b_sb_m}$, a sufficient condition for a finite lower bound is
\begin{equation}\label{stable_sector}
\kappa\geq0,\qquad\kappa_\psi\geq0.
\end{equation}
On a bounded domain with periodic boundary conditions, a square-integrable pinning field and fixed external gauge potential, the positive quartic terms control the lower-order terms, while Eq.~\eqref{stable_sector} makes the gradient energy nonnegative. The resulting functional has a finite lower bound. The proof and the status of negative gradient couplings are given in Appendix~\ref{app:stability}.

If either mixed-gradient coefficient is negative, positivity evaluated only at the homogeneous stationary amplitudes is insufficient for global stability. With unrestricted amplitudes, a sufficiently large constant background of the other field makes the corresponding gradient coefficient negative; arbitrarily short-wavelength, small-amplitude oscillations then drive the truncated continuum energy downward. Negative couplings can instead be studied in a specified bounded-amplitude effective sector, or in an explicitly stabilized extended functional.

A lattice cutoff provides another regularization. For the square-lattice bond form in Appendix~\ref{app:lattice}, $\kappa\geq0$ and $\kappa_\psi<0$, a sufficient coercive bound is
\[
8|\kappa_\psi|+\max(-\gamma,0)<\sqrt{b_sb_m},
\]
whereas a staggered condensate requires $\gamma+8\kappa_\psi\geq-\sqrt{b_sb_m}$ as a necessary condition. Appendix~\ref{app:stability} gives the derivation. The negative-$\kappa_\psi$ results below satisfy the sufficient bound; they are cutoff-dependent lattice results, not stable realizations of the unrestricted continuum functional.

\begin{figure}[tb]\centering\includegraphics[width=.85\linewidth]{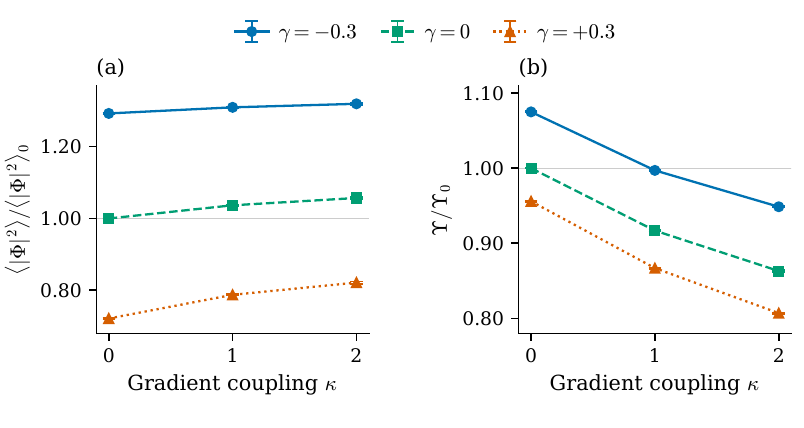}\caption{Monte Carlo response ratios at $T=0.5\approx0.64\,T_{c0}$ and $\kappa_\psi=0$, reproduced from Table~II of Ref.~\cite{mel2025} (model in Appendix~\ref{app:lattice}). (a) Mean-square envelope amplitude and (b) helicity modulus, each relative to the reference model ($\gamma=\kappa=\kappa_\psi=0$) at the same $T$ and $W$. Lines join points at fixed $\gamma$. Error bars show pointwise one-standard-error uncertainties across independent chains (three for $(\gamma,\kappa)=(+0.3,2)$ and six otherwise), propagating numerator and reference errors in quadrature; the reference self-ratio at $(0,0)$ is identically one. Shared reference errors correlate the other ratios, and bars can be smaller than markers. History-dependent systematic uncertainty, up to $1.9\%$ in (b), is not included.}\label{fig:mc}\end{figure}

\paragraph{Numerical provenance.}
Figure~\ref{fig:mc} and Table~\ref{tab:mc_provenance} reproduce the central estimates of Table~II of the second version of Ref.~\cite{mel2025}. The calculation uses the lattice discretization of Eq.~\eqref{F}, with parameters, pinning statistics and updates specified in Appendix~\ref{app:lattice}. Energies are measured in $\varepsilon_0$, with $k_B=1$; the reference model has $\gamma=\kappa=\kappa_\psi=0$ at the same temperature and pinning strength. Its transition is at $T_{c0}=0.785$, so the plotted temperature $T=0.5\approx0.64\,T_{c0}$ lies below the reference transition.

\begin{table}[tb]\centering\small
\caption{Data of Fig.~\ref{fig:mc}, from Table~II of Ref.~\cite{mel2025}, at $T=0.5$ and $\kappa_\psi=0$, relative to the reference model at the same $T$ and $W$.}\label{tab:mc_provenance}
\begin{tabular}{@{}rrrrr@{}}\toprule
$\gamma$ & $\kappa$ & $\langle|\Phi|^2\rangle$ ratio & $\Upsilon$ ratio & $n_{\rm chains}$\\\midrule
$-0.3$&0&1.29&1.07&6\\
$-0.3$&1&1.31&1.00&6\\
$-0.3$&2&1.32&0.95&6\\
$0$&1&1.04&0.92&6\\
$0$&2&1.06&0.86&6\\
$+0.3$&0&0.72&0.96&6\\
$+0.3$&1&0.79&0.87&6\\
$+0.3$&2&0.82&0.81&3\\\bottomrule
\end{tabular}\end{table}

\section{Phase Rigidity and Correlation Functions}
\subsection{Phase-Only Approximation}
For fixed amplitudes $A$ and $S$, the phase-dependent part of Eq.~\eqref{F}, including the quenched pinning term, is
\begin{align}\label{Fphase}
F_{\rm phase}=\int\dd^2r\,\big[&(K_m+\kappa S^2)A^2|\nabla\varphi|^2\nonumber\\
&+(K_s+\kappa_\psi A^2)S^2|\nabla\theta-q_s\mathbf A/\hbar|^2\nonumber\\
&-A\,\operatorname{Re}(h^*e^{i\varphi})\big].
\end{align}
The pinning contribution retains its dependence on the translational phase even at fixed amplitude. The gradient coefficients in Eq.~\eqref{Fphase} precede the disorder and thermal averages used to obtain a thermodynamic helicity modulus.

For a uniform system with fixed amplitudes and no fluctuations, a twist $\theta\mapsto\theta+\delta x/L_x$ gives a free-energy increment $\Delta F=(K_s+\kappa_\psi A^2)S^2(\delta/L_x)^2|\Omega|$. If the helicity modulus is defined as $\Upsilon=(L_x^2/|\Omega|)\partial_\delta^2F|_{\delta=0}$, then
\begin{equation}\label{Ymf}\Upsilon_{\rm bare}=2(K_s+\kappa_\psi A^2)S^2.\end{equation}
In a fluctuating model, $\Upsilon$ must instead be obtained by differentiating the full thermodynamic free energy.

\subsection{Helicity Modulus}\label{sec:helicity}
Let $Z(\delta)=\int\mathcal D\psi\mathcal D\Phi\,\exp[-\beta F_\delta]$, with $\beta=1/k_BT$, for a boundary twist $\delta$ applied to the superconducting field. Define $\mathcal F(\delta)=-\beta^{-1}\log Z(\delta)$. The helicity modulus is \cite{fisher1973}
\begin{equation}\label{helicity}\Upsilon=\frac{L_x^2}{|\Omega|}\left.\frac{\partial^2\mathcal F}{\partial\delta^2}\right|_0
=\frac{L_x^2}{|\Omega|}\left[\left\langle F_\delta''\right\rangle-\beta\left(\left\langle(F_\delta')^2\right\rangle-\left\langle F_\delta'\right\rangle^2\right)\right]_0.
\end{equation}
Primes denote derivatives with respect to the twist. The second, fluctuation term is generally nonzero. In a lattice simulation the twist derivative must be applied to the same bond weights used in the Hamiltonian; retaining $K_s$ while omitting the $\kappa_\psi|\Phi|^2$ contribution from the estimator would be inconsistent.

\begin{figure}[ht]\centering\includegraphics[width=.70\linewidth]{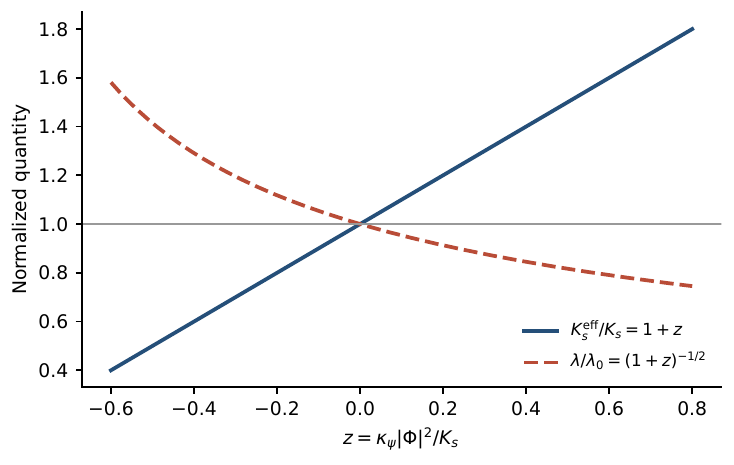}\caption{Exact fixed-amplitude algebra for $z=\kappa_\psi A^2/K_s$: the bare coefficient ratio $1+z$ and the corresponding conditional London penetration-depth ratio $(1+z)^{-1/2}$. For $z<0$ the continuum functional is unbounded below (Appendix~\ref{app:stability}); that branch requires a lattice cutoff or another explicit stabilization.}\label{fig:stiff}\end{figure}

Figure~\ref{fig:stiff} illustrates the local London conversion. With geometry and other factors held fixed, $\lambda^{-2}\propto\Upsilon$ and hence
\begin{equation}\label{lambda}\frac{\lambda}{\lambda_0}=\left(\frac{\Upsilon}{\Upsilon_0}\right)^{-1/2}.\end{equation}
A 10\% change in stiffness is not a 10\% change in penetration depth. Equation~\eqref{lambda} also does not replace a microscopic current-response calculation or the treatment of interlayer electrodynamics.

\subsection{Amplitude and Phase Correlation Functions}
The phase-only and amplitude-weighted correlations are
\begin{align}\label{corr}
C_{\varphi}(\mathbf r)&=\left\langle e^{i[\varphi(\mathbf r)-\varphi(\mathbf 0)]}\right\rangle,\\
C_{\Phi}(\mathbf r)&=\left\langle A(\mathbf r)A(\mathbf 0)e^{i[\varphi(\mathbf r)-\varphi(\mathbf 0)]}\right\rangle.
\end{align}
They are generally unequal when the amplitude is spatially heterogeneous. The Fourier transform of $C_\Phi$ enters a model structure factor, while $C_\varphi$ characterizes phase organization after the amplitude has been divided out.

When $A$ becomes small, phase extraction becomes ill-conditioned. Any operational phase map must therefore specify an amplitude mask or uncertainty model. For short-range textures, spatial averaging, finite field of view, and disorder pinning alter both fitted lengths. A quantitative length requires an estimator whose bias has been tested on known synthetic fields and whose uncertainty accounts for correlated spatial samples.

\subsection{Structure Factors and Correlation Lengths}
A complex envelope has an amplitude-weighted two-point function $C_\Phi(\mathbf r)=\langle\Phi(\mathbf r)\Phi^*(\mathbf0)\rangle$ and a phase-only correlator $C_\varphi(\mathbf r)=\langle e^{i[\varphi(\mathbf r)-\varphi(\mathbf0)]}\rangle$ where the phase is defined. They are not interchangeable because
\begin{equation}
C_\Phi(\mathbf r)=\langle A(\mathbf r)A(\mathbf0)e^{i[\varphi(\mathbf r)-\varphi(\mathbf0)]}\rangle.
\end{equation}
Only when amplitude fluctuations can be factorized from phase fluctuations does $C_\Phi$ reduce to an amplitude prefactor times $C_\varphi$. Disorder and coupling to $|\psi|^2$ generally spoil such factorization. The envelope structure factor is $S_\Phi(\mathbf k)=\int d^2r\,e^{-i\mathbf k\cdot\mathbf r}C_\Phi(\mathbf r)$; the physical density contribution is shifted to the neighbourhoods of $\pm\mathbf Q$, weighted by the form factor and higher harmonics. A scattering width consequently constrains an amplitude-weighted spatial correlation.

For a finite numerical lattice, a second-moment estimator may be constructed from $S_\Phi(\mathbf0)$ and $S_\Phi(\mathbf k_{\min})$, provided the relevant structure factor is resolved and positive. Exponential-fit lengths are also useful but depend on the fit range and the maximum available separation. When a fitted length approaches the box size, quoting additional significant figures does not establish convergence. The numerical comparison in Fig.~\ref{fig:mc} is restricted to amplitude and helicity-modulus ratios, which are separate from fitted spatial correlation lengths.

\begin{table}[!htbp]\centering\small
\caption{Correlation and rigidity notation used in the coherence analysis.}\label{tab:coherence_symbols}
\begin{tabular}{@{}p{.23\linewidth}p{.67\linewidth}@{}}\toprule
Symbol & Definition and convention\\\midrule
$\Upsilon$ & Thermodynamic helicity modulus defined by a uniform boundary twist.\\
$\delta,L_x,L_y$ & Imposed total twist and sample dimensions along the two in-plane axes.\\
$\mathcal F$ & Twist-dependent thermodynamic free energy $-\beta^{-1}\log Z(\delta)$.\\
$C_\varphi(r)$ & Phase-only translational correlation of the envelope.\\
$C_\Phi(r)$ & Amplitude-weighted complex-envelope correlation.\\
$\xi_\varphi,\xi_\Phi$ & Lengths extracted from separately specified phase-only and full-envelope estimators.\\
$S_\Phi(\mathbf k)$ & Fourier-space structure factor of the electronic envelope.\\
$\lambda,\lambda_0$ & Penetration depth and reference value in a fixed electrodynamic geometry.\\
\bottomrule\end{tabular}\end{table}

An experimental correlation length is always conditional on the measured channel and the estimator used. A phase-only length weights low- and high-amplitude regions differently from a complex-envelope structure factor. This distinction is essential when a numerical length is compared with a diffraction linewidth or a spatially filtered tunnelling map.

\section{Analytical Response of the Coupled Fields}
\subsection{Uniform Amplitude Response}
For fixed uniform superconducting amplitude $S$, the electronic sector has the effective quadratic coefficient $a_m+\gamma S^2$. In the ordered-amplitude regime,
\begin{equation}\label{A2}A^2=-\frac{a_m+\gamma S^2}{b_m},\qquad a_m+\gamma S^2<0.\end{equation}
At fixed $S$ and fixed $b_m$, $\partial A^2/\partial\gamma=-S^2/b_m$. A positive $\gamma$ therefore reduces $A^2$ in this restricted setting. This derivative is not a statement about the total temperature derivative of $A^2$ when $S$, the disorder response, or the other couplings change self-consistently.

Likewise, for fixed $A$ and in the absence of gradients, $S^2=-(a_s+\gamma A^2)/b_s$ when the right-hand side is positive. The mutual feedback is summarized by the matrix equation~\eqref{matrix}. These elementary results already show why the sign of a local amplitude correlation cannot be read from $\gamma$ without specifying which degrees of freedom fluctuate and what is held fixed.

The selected numerical results in Fig.~\ref{fig:mc} illustrate why a modulation-intensity response cannot be substituted for a stiffness response. For $\gamma=-0.3$ and $\kappa=0,1,2$, the amplitude ratios are $1.29,1.31,1.32$, whereas the helicity ratios are $1.07,1.00,0.95$. For $\gamma=+0.3$ the respective sequences are $0.72,0.79,0.82$ and $0.96,0.87,0.81$. These are fixed-temperature comparisons between different model couplings, not an experimental evolution through $T_c$. We do not use the corresponding finite-size-sensitive phase lengths to establish a numerical coherence enhancement.

\subsection{Collective Amplitude Eigenmodes and Reduced Susceptibility}
\begin{figure}[tb]\centering
\includegraphics[width=.73\linewidth]{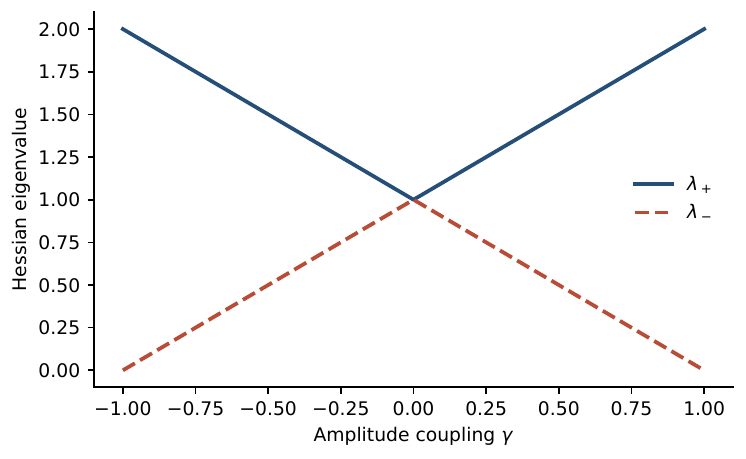}
\caption{Exact eigenvalues $\lambda_\pm$ of the uniform amplitude-response Hessian for $b_s=b_m=1$ and variable $\gamma$. The vertical axis gives dimensionless Hessian eigenvalues; the dashed line is the softer eigenvalue $1-|\gamma|$. The vanishing eigenvalue at $|\gamma|=1$ marks the boundary of local interior stability in this parameterization.}\label{fig:eigenmodes}
\end{figure}

The uniform stationary equations in Eq.~\eqref{matrix} define a symmetric amplitude-response matrix
\begin{equation}\label{Hamp}
B=\begin{pmatrix}b_s&\gamma\\\gamma&b_m\end{pmatrix},\qquad
\lambda_{\pm}=\frac{b_s+b_m\pm\sqrt{(b_s-b_m)^2+4\gamma^2}}{2}.
\end{equation}
Figure~\ref{fig:eigenmodes} displays the symmetric special case of these eigenvalues. The eigenvalues of the source-defined susceptibility $B^{-1}$ are $\lambda_{\pm}^{-1}$. The eigenvectors represent combinations of squared condensate and modulation amplitudes rather than separate microscopic normal modes. In the interior coexistence region, $b_s>0$ and $D=b_sb_m-\gamma^2>0$ ensure $\lambda_{\pm}>0$. The smaller eigenvalue controls the most responsive combination of amplitudes. In particular, for $b_s=b_m=b$, the two eigenvalues reduce to $b\pm|\gamma|$, and the susceptibility in the soft direction increases as $(b-|\gamma|)^{-1}$.

A complementary derivation eliminates the condensate-amplitude variable $x$ at fixed $y$ by solving $\partial f_0/\partial x=0$:
\begin{equation}\label{schur}
x_*(y)=-\frac{a_s+\gamma y}{b_s},\qquad
f_{\mathrm{red}}(y)=a_my+\frac{b_m}{2}y^2-\frac{(a_s+\gamma y)^2}{2b_s},
\end{equation}
where $x_*(y)>0$ is required for the interior branch. Its effective curvature is
\begin{equation}\label{beff}
\frac{\dd^2 f_{\mathrm{red}}}{\dd y^2}=b_m-\frac{\gamma^2}{b_s}=\frac{D}{b_s}.
\end{equation}
Equations~\eqref{Hamp}--\eqref{beff} establish the same local stability threshold by two independent algebraic constructions. The reduction also shows how allowing one amplitude to relax increases the susceptibility of the other relative to a calculation with that amplitude held fixed. The sign of the off-diagonal susceptibility is $-\gamma/D$, while the eigenvalues depend on $\gamma^2$; these are different pieces of information about the coupled response.

\begin{table}[ht]\centering\small
\caption{Quantities in the collective-amplitude response.}\label{tab:mode_symbols}
\begin{tabular}{@{}p{.24\linewidth}p{.64\linewidth}@{}}\toprule
Symbol & Definition \\\midrule
$B$ & Hessian of $f_0$ with respect to $x=|\psi|^2$ and $y=|\Phi|^2$.\\
$\lambda_{\pm}$ & Eigenvalues of $B$, ordered $\lambda_+\geq\lambda_-$.\\
$f_{\rm red}$ & Uniform free-energy density after stationary elimination of $x$.\\
$D$ & Determinant $b_sb_m-\gamma^2$; $D>0$ for an interior minimum.\\
\bottomrule\end{tabular}\end{table}

\subsection{Disorder-Dependent Coherence Scales}
Under a slowly varying amplitude approximation, the phason-gradient coefficient is $\rho_{\varphi}=2(K_m+\kappa S^2)A^2$. For a random-field texture, a balance between gradient energy and pinning energy can yield a characteristic domain scale \cite{imryma1975}. The exponent of this balance depends on dimensionality, pinning statistics, boundary conditions and whether the amplitude is itself soft.

For a quenched random field, the phase correlation length is defined operationally by the decay of the disorder-averaged phase correlator. Its dependence on the pinning variance and the gradient coefficient follows from the full statistical ensemble rather than a universal single-power scaling. The amplitude-weighted correlation length and the phase-only length are consequently reported as distinct quantities.

\subsection{Coupling Dependence of the Helicity Modulus}
The coefficient $K_s^{\rm eff}$ rises when $\kappa_\psi>0$ at fixed $A$. Nevertheless, $\Upsilon$ in Eq.~\eqref{helicity} can decrease if the condensate amplitude is suppressed or phase fluctuations become stronger. The $\kappa$ term also changes the superconducting configuration through the $\kappa|\nabla\Phi|^2\psi$ term in Eq.~\eqref{ELs}. Therefore the sign of $\kappa$, or even the sign of $\kappa_\psi$, does not alone fix the sign of the final thermodynamic response.

\paragraph{Condensate-gradient scan.}
As a further numerical result of this study, Table~\ref{tab:kpsi} gives the $\kappa_\psi$ sensitivity at fixed $\kappa=1$, using the same lattice model and bare parameters. Here each coupled stiffness is divided by the reference with $\gamma=\kappa=0$ at the same $\kappa_\psi$, rather than by the $\kappa_\psi=0$ reference of Fig.~\ref{fig:mc}. For this normalization Eq.~\eqref{main_enhancement} contains the factor $(1+\kappa_\psi y/K_s)/(1+\kappa_\psi y_0/K_s)$, with $y,y_0$ from Eq.~\eqref{solution}. It predicts the signs of all four shifts away from the central column: negative $\kappa_\psi$ lowers the cooperative ratio and raises the competing ratio, whereas positive $\kappa_\psi$ reverses these changes. The reference stiffness itself changes to $0.82$ and $1.20$ of its $\kappa_\psi=0$ value, compared with bare factors $0.85$ and $1.25$. The negative-coupling column is a cutoff-dependent lattice calculation satisfying the bound in Sec.~4.2.

\begin{table}[ht]\centering\small
\caption{Helicity-modulus ratios at $T=0.5$ and $\kappa=1$, divided by the reference at the same $\kappa_\psi$. There are three chains per point in the outer columns and six in the central column.}\label{tab:kpsi}
\begin{tabular}{@{}rrrr@{}}\toprule
 & \multicolumn{3}{c}{$\kappa_\psi$}\\
$\gamma$ & $-0.30$ & $0$ & $+0.50$\\\midrule
$-0.3$ & 0.95 & 1.00 & 1.06\\
$+0.3$ & 0.88 & 0.87 & 0.84\\
\bottomrule\end{tabular}\end{table}

\begin{table}[ht]\centering\caption{Distinct roles of the retained couplings. The last column states what cannot be inferred from the bare coefficient alone.}\label{tab:couplings}\small\begin{tabular}{@{}p{.13\linewidth}p{.35\linewidth}p{.41\linewidth}@{}}\toprule
Term & Direct contribution & Not determined without solving the coupled state\\\midrule
$\gamma I_0$ & Uniform amplitude masses & Total stiffness or phase-domain length\\
$\kappa I_1$ & Electronic gradient coefficient; condensate mass feedback & Net phase coherence or helicity modulus\\
$\kappa_\psi I_2$ & Condensate gradient coefficient; envelope mass feedback & Net penetration depth or transition temperature\\\bottomrule\end{tabular}\end{table}

\subsection{Linear Response of the Coexistence State}
A stationary uniform solution does not specify how the two fields respond to weak perturbations. To obtain a local amplitude susceptibility, write $x=S^2$ and $y=A^2$ and perturb the quadratic coefficients by $a_s\mapsto a_s+\delta a_s$ and $a_m\mapsto a_m+\delta a_m$. The stationarity conditions are $a_s+b_s x+\gamma y=0$ and $a_m+b_m y+\gamma x=0$. Linearization yields
\begin{equation}\label{responsematrix}
\begin{pmatrix}b_s&\gamma\\\gamma&b_m\end{pmatrix}
\begin{pmatrix}\delta x\\\delta y\end{pmatrix}=-\begin{pmatrix}\delta a_s\\\delta a_m\end{pmatrix}.
\end{equation}
For $D=b_sb_m-\gamma^2>0$, the susceptibility is therefore
\begin{equation}\label{responseinverse}
\begin{pmatrix}\delta x\\\delta y\end{pmatrix}
=-\frac1D\begin{pmatrix}b_m&-\gamma\\-\gamma&b_s\end{pmatrix}
\begin{pmatrix}\delta a_s\\\delta a_m\end{pmatrix}.
\end{equation}
The inverse Hessian has off-diagonal element $(B^{-1})_{sm}=-\gamma/D$. For perturbations of the quadratic coefficient $a_m$, however, the actual derivative is $\partial x/\partial a_m=+\gamma/D$ because $\delta\mathbf u=-B^{-1}\delta\mathbf a$. Defining an external source $\mathbf j=-\mathbf a$ instead gives $\partial\mathbf u/\partial\mathbf j=B^{-1}$. These distinct source conventions determine the corresponding cross-response signs. The divergence of the linear response as $D\to0^+$ signals the loss of invertibility of the uniform stationary equations, but does not alone determine the nature of the transition once gradients, fluctuations and higher-order operators are included.

In a spatially varying setting, $\delta a_s(\mathbf r)$ and $\delta a_m(\mathbf r)$ may represent external tuning or a coarse-grained structural environment. The response is then nonlocal, and its kernel is the inverse of a differential operator rather than the constant matrix in Eq.~\eqref{responsematrix}. For small fluctuations with wave vector $\mathbf k$, the diagonal amplitude-gradient coefficients add terms proportional to $k^2$. The resulting susceptibility depends on the same couplings but also on the background amplitudes and the boundary conditions. A positive local cross-response is consequently not equivalent to a positive experimental correlation between two spectral maps: the latter includes the response kernels and common sources of spatial variation.

For a uniform background with $x>0$ and $y>0$, a fixed-phase amplitude fluctuation has $|\nabla\psi|^2=(\nabla x)^2/(4x)$ and $|\nabla\Phi|^2=(\nabla y)^2/(4y)$ to quadratic order. The momentum-resolved amplitude Hessian of Eq.~\eqref{F} is therefore
\begin{equation}\label{finitek}
B(\mathbf k)=\begin{pmatrix}
b_s+\dfrac{K_s+\kappa_\psi y}{2x}|\mathbf k|^2 & \gamma\\[0.4em]
\gamma & b_m+\dfrac{K_m+\kappa x}{2y}|\mathbf k|^2
\end{pmatrix}.
\end{equation}
The linear response to spatially modulated quadratic coefficients is $\delta\mathbf u(\mathbf k)=-B(\mathbf k)^{-1}\delta\mathbf a(\mathbf k)$, with $\mathbf u=(x,y)^T$ and $\mathbf a=(a_s,a_m)^T$. In particular,
\begin{equation}\label{crossfinite}
[B(\mathbf k)^{-1}]_{sm}=-\frac{\gamma}{\det B(\mathbf k)}.
\end{equation}
These relations provide an explicit finite-wave-vector response of the coupled amplitude sector. They are obtained for a uniform, phase-fixed reference configuration, so they describe amplitude fluctuations rather than the separate translational-phase or superconducting phase modes. For an interior minimum with $D>0$ and strict inequalities in Eq.~\eqref{Keff}, $\det B(\mathbf k)\geq D>0$ for all $\mathbf k$, so the cross-susceptibility has the sign of $-\gamma$ at every wave vector; the gradient couplings set its spatial range.

The site-wise Pearson coefficient between $|\psi_i|$ and $|\Phi_i|$ at $T=0.5$, listed in Table~II of Ref.~\cite{mel2025}, is $+0.11,+0.05,-0.04,-0.06$ for $\gamma=-0.6,-0.3,+0.3,+0.6$ at $\kappa=\kappa_\psi=0$, and its magnitude is below $0.002$ at $\gamma=0$, $\kappa=1,2$. Thus its sign follows $-\gamma$, in agreement with the linearized amplitude response in Eq.~\eqref{crossfinite}. The recorded estimator uses moduli rather than their squares; around positive uniform amplitudes, $\delta|\psi|^2\simeq2S\,\delta|\psi|$ and $\delta|\Phi|^2\simeq2A\,\delta|\Phi|$, so the two Pearson coefficients coincide to linear order, but need not agree in the full fluctuating ensemble.

\begin{table}[!htbp]\centering\small
\caption{Response notation in the uniform linearization.}\label{tab:response_symbols}
\begin{tabular}{@{}p{.23\linewidth}p{.67\linewidth}@{}}\toprule
Symbol & Definition and convention\\\midrule
$\delta a_s,\delta a_m$ & Small changes in the two quadratic coefficients.\\
$\delta x,\delta y$ & Induced changes in the squared field magnitudes.\\
$D=b_sb_m-\gamma^2$ & Determinant controlling invertibility of the uniform linear response.\\
$\mathbf k$ & Wave vector of a spatial perturbation when gradient terms are retained.\\
$B(\mathbf k)$ & Fixed-phase amplitude Hessian including the leading spatial-gradient response.\\
\bottomrule\end{tabular}\end{table}

The matrix response distinguishes an externally imposed perturbation from the field that responds to it. The source-defined off-diagonal susceptibility depends on the amplitude coupling $\gamma$, but the sign of a measured spatial correlation need not coincide with this response: disorder, tunnelling matrix elements and the simultaneous action of other couplings can alter the observed covariance.

\subsection{Local Wave-Vector Representation}
The decomposition in Eq.~\eqref{density} is invariant under a change of the reference carrier provided the envelope transforms oppositely. Let $\mathbf Q'=\mathbf Q+\boldsymbol\delta$ and $\Phi'(\mathbf r)=\Phi(\mathbf r)e^{-i\boldsymbol\delta\cdot\mathbf r}$. Then $\Phi'e^{i\mathbf Q'\cdot\mathbf r}=\Phi e^{i\mathbf Q\cdot\mathbf r}$ exactly. Where the phase is smooth, $\mathbf Q'+\nabla\arg\Phi'=\mathbf Q+\nabla\arg\Phi$. Thus the physical local wave vector is invariant under this carrier convention, while the envelope gradient by itself is not. A comparison of phase-gradient energies across different carriers must respect this transformation. In particular, with $\mathbf Q_0\mapsto\mathbf Q_0+\boldsymbol\delta$, the equivalent envelope operator is $\nabla\mapsto\nabla+i\boldsymbol\delta$ acting on the transformed envelope: $(\nabla+i\boldsymbol\delta)\Phi'=e^{-i\boldsymbol\delta\cdot\mathbf r}\nabla\Phi$. The ordinary gradient form is therefore tied to the fixed stationary-carrier convention used in Eq.~\eqref{qkernel}.

A finite-window demodulation does not reconstruct an arbitrary $\Phi$ without assumptions. Window size trades spatial localization against wave-vector resolution; the phase becomes unstable where the recovered amplitude approaches the noise floor. A form-factor projection also requires a registration of the sublattice geometry before a bond-centred signal can be interpreted. These are consequences of the observation operator, not adjustable physical couplings of the free energy.

\section{Spectroscopic Response Functions}
\subsection{Local Tunnelling Response}
A scanning tunnelling spectrometer measures a differential conductance $g(\mathbf r,E)$ at bias energy $E=eV$ which, under restrictive tunnelling-matrix-element and thermal conditions, is related to the local electronic spectral density. The measured conductance is not the order parameter $\Phi$. A linearized measurement model at energy $E$ may be written
\begin{equation}\label{STS}g(\mathbf r,E)=g_0(\mathbf r,E)+\Ree\{M_E(\mathbf r)\Phi(\mathbf r)e^{i\mathbf Q\cdot\mathbf r}\}+\eta(\mathbf r,E),\end{equation}
where $M_E$ is an energy-dependent response and $\eta$ contains measurement error and unmodelled contributions. The form-factor projection and atomic registration are included in the definition of the response channel. Even when a local complex conductance modulation can be extracted, its amplitude depends on $M_E$ and its phase can be shifted by the complex response.

A practical test can therefore compare the reproducibility of spatial patterns across energy, acquisition direction and independently registered fields. Quantities of interest include a local amplitude distribution, a wave-vector distribution, the amplitude-weighted spatial correlation and a phase-coherence estimator with stated masking.

A gap extracted from the peak location of $g(\mathbf r,E)$ is an energy scale, not $S(\mathbf r)$ itself. Likewise, a quasiparticle-interference extinction scale and a pseudogap scale are not interchangeable definitions. A correlation between an STS gap and the electronic modulation can arise from common structural or doping inhomogeneity; it is not by itself a measurement of $\gamma$. Distinguishing superconducting spectral coherence from a gap magnitude is therefore part of the experimental inference problem.

\subsection{Momentum-Resolved Spectral Function}
ARPES is sensitive, modulo matrix elements and occupation factors, to the single-particle spectral function. A static finite-$\mathbf Q$ potential can mix states at $\mathbf k$ and $\mathbf k+\mathbf Q$. To exhibit the minimum additional microscopic input, consider the two-band Hamiltonian
\begin{equation}\label{Hk}H_{\mathbf k}=\begin{pmatrix}\epsilon_{\mathbf k}&V_{\mathbf Q}\\V_{\mathbf Q}^*&\epsilon_{\mathbf k+\mathbf Q}\end{pmatrix},\qquad V_{\mathbf Q}=g_m\Phi.\end{equation}
Its eigenvalues are
\begin{equation}\label{Ek}E_\pm=\frac{\epsilon_{\mathbf k}+\epsilon_{\mathbf k+\mathbf Q}}2\pm\sqrt{\left(\frac{\epsilon_{\mathbf k}-\epsilon_{\mathbf k+\mathbf Q}}2\right)^2+|V_{\mathbf Q}|^2}.\end{equation}
The avoided crossing is an exact consequence of this two-band Hamiltonian, not of the GL functional alone. For finite domains, disorder and dynamic fluctuations, the spectral weight depends on a self-energy or disorder average and cannot be inferred from Eq.~\eqref{Ek} by merely inserting a short correlation length.

In an idealized quasistatic limit, the ARPES intensity has the form\begin{equation}I(\mathbf k,\omega)\propto |M_{\rm ARPES}|^2f(\omega)\mathcal A(\mathbf k,\omega),\end{equation} with a material-dependent matrix element. The spectral function $\mathcal A$ requires a Green function, an electronic dispersion and a scattering model. Thus momentum-space reconstruction is a possible diagnostic of electronic modulation under additional assumptions.

\subsection{Spectral Weights of the Two-Band Model}
The eigenvalues in Eq.~\eqref{Ek} alone do not determine the spectral visibility of the reconstructed branches. Writing $\Delta\epsilon=\epsilon_{\mathbf k}-\epsilon_{\mathbf k+\mathbf Q}$, the probability weights in the original $\mathbf k$ component are
\begin{equation}\label{weights}
w_\pm(\mathbf k)=\frac12\left[1\pm\frac{\Delta\epsilon}{\sqrt{(\Delta\epsilon)^2+4|V_{\mathbf Q}|^2}}\right].
\end{equation}
At a degeneracy the two branches share the original-state weight equally, while away from the crossing one branch approaches the unperturbed state. These are exact statements for the static two-band Hamiltonian. In a fluctuating short-range texture, finite-domain averaging and the electronic self-energy determine whether the reconstructed branches remain spectroscopically distinguishable.

\begin{table}[!htbp]\centering\small
\caption{Symbols in the illustrative tunnelling and photoemission forward models.}\label{tab:spectro_symbols}
\begin{tabular}{@{}p{.23\linewidth}p{.67\linewidth}@{}}\toprule
Symbol & Definition and convention\\\midrule
$g(\mathbf r,E),g_0$ & Differential conductance and its unmodulated contribution at energy $E$.\\
$M_E,\eta$ & Complex energy-dependent tunnelling response and unmodelled/noise contribution.\\
$\epsilon_{\mathbf k}$ & Normal-state electronic dispersion at crystal momentum $\mathbf k$.\\
$V_{\mathbf Q},g_m$ & Modulation-induced electronic mixing potential and its microscopic coupling.\\
$E_\pm,w_\pm$ & Eigenenergies and original-momentum spectral weights of the two-state model.\\
$I(\mathbf k,\omega)$ & Photoemission intensity at momentum and frequency.\\
$\mathcal A,f(\omega),M_{\rm ARPES}$ & Spectral function, Fermi occupation and photoemission matrix element.\\
\bottomrule\end{tabular}\end{table}

These expressions provide a route from an electronic texture to candidate observables, not an assertion that the phenomenological free energy alone predicts a measured spectrum. In STM/STS the transfer function $M_E$ may depend on energy and local environment. In ARPES the spectral function must incorporate finite lifetime, matrix elements and the distribution of modulation domains before the static reconstruction picture can be confronted with data.

\subsection{Superfluid Stiffness}
The most direct comparison is with a thermodynamic or electrodynamic stiffness measurement independent of single-particle spectral features. The relation between the model helicity modulus and a measured penetration depth depends on layer thickness, dimensionality, units and fluctuation effects. We therefore keep $\Upsilon$ as the primary theoretical observable and introduce the penetration depth only after specifying a separate electrodynamic conversion. An experimental programme need not determine all couplings individually to constrain their combinations.

\begin{table}[ht]\centering\caption{Observation map and principal inference limitation.}\label{tab:obs}\small\begin{tabular}{@{}p{.21\linewidth}p{.34\linewidth}p{.35\linewidth}@{}}\toprule
Technique & Candidate observable & Required additional assumption\\\midrule
STM/STS & Spatial conductance modulation and energy dependence & Tunnelling response, registration, form factor\\
ARPES & Spectral reconstruction and weight redistribution & Electronic Hamiltonian, matrix elements, self-energy\\
Penetration depth & Superfluid response & Electrodynamic geometry and stiffness conversion\\
Scattering & Envelope-weighted structure factor & Scattering weights, resolution and background\\\bottomrule\end{tabular}\end{table}

\section{Discussion}
\subsection{Relation to the Earlier MEL Enhancement Regime}
The first version of Ref.~\cite{mel2025} used a coupling linear in the modulation density. For incommensurate $\mathbf Q$ that term is not translation invariant and is replaced here by $|\Phi|^2|\psi|^2$. Its enhancement window is not used quantitatively: the present comparison instead uses Eq.~\eqref{main_enhancement} and the Monte Carlo response in Fig.~\ref{fig:mc}. At $\kappa_\psi=0$, the homogeneous stiffness rises for $\gamma<0$, together with the modulation amplitude; the $\kappa=0$ Monte Carlo ratios follow the same sign rule. The gradient coupling $\kappa$, absent from the homogeneous reduction, lowers the stiffness at each plotted $\gamma$.

The phenomenological framework describes the coupled response of electronic modulation and superconducting coherence at the level of coarse-grained fields. Its usefulness is that the same parameters enter distinct response functions. In a disordered system a peak width depends on pinning and the estimator; the finite-size behaviour of the phase length must be established before using it as a quantitative target \cite{orgad2024,caplan2017,delmaestro2006,robertson2006}. The related literature on phonon anomalies, vortex electrodynamics and penetration depth \cite{pumpin1990} supplies physical context but not a direct determination of the three couplings \cite{reznik2006,letacon2014,sonier1994,sonier2007,blatter1994}.\par

The model separates three effects that are often combined in an informal statement that electronic order ``competes with'' or ``reinforces'' superconductivity. A local amplitude coupling changes the equilibrium magnitudes of the two fields; a modulation-gradient coupling changes the energetic cost of electronic texture; and a condensate-gradient coupling changes the local cost of superconducting phase variation. Their feedback is simultaneous. A sign assignment to one coefficient cannot be substituted for the measured response of the fully fluctuating system.

The distinction between $C_\varphi$ and $C_\Phi$ matters as much. A short-range texture may have a large local amplitude but a rapidly varying phase. A single global Fourier peak averages over this organization. Conversely, a local demodulation algorithm can manufacture an apparent length scale through its window even when the underlying signal is weak. The formulation therefore motivates analysis of spatial organization, but does not validate any particular demodulation method by itself.

Symmetry admissibility and experimental identification are separate questions: a coupling allowed by symmetry need not be large in a given material.

The temperature dependence of the preferred wave vector is controlled by the mixed derivative of the effective kernel in Eq.~\eqref{qderivative}. In the fixed-carrier local truncation a constant amplitude coupling does not move the quadratic minimum. A shift linear in $S^2$ follows from the symmetry-allowed first-order invariant of Sec.~2.2, whose coefficient is not fixed by symmetry; an intrinsic temperature dependence of the bare kernel adds to it.

The single-$\mathbf Q$ sector applies to a selected modulation orientation, and its gradient expansion is organized for envelope variations slow relative to the carrier. The continuum stability conditions follow from the effective gradient coefficients; higher-order operators specify the extension when those coefficients approach zero. The analytical figures represent stationary solutions and minimal spectral constructions, while the numerical response ratios are identified separately.

Amplitude, phase-correlation and helicity responses can therefore be compared within one model. For disordered numerical ensembles, disorder averaging, autocorrelation times and finite-size scaling enter the estimators. Comparison with scattering is obtained by applying the relevant measurement kernel to the simulated electronic field.

\section{Conclusion}
A finite-wave-vector electronic envelope and a superconducting condensate admit distinct local-amplitude and mixed-gradient couplings. Their variation yields a coupled set of equations in which amplitude response, translational coherence and superconducting phase rigidity are related but nonidentical. Uniform coexistence conditions, collective-amplitude eigenvalues, the reduced susceptibility and gradient-stability restrictions follow analytically; the helicity modulus requires a thermodynamic twist response rather than inspection of a bare coefficient. Spatial spectroscopy can interrogate the organization of the electronic texture, while photoemission probes its consequences for single-particle spectra only after a microscopic coupling is specified. Each of these relations can be tested once the corresponding measurement model is specified.

\section*{Acknowledgements}
We are grateful to colleagues involved in the experimental MEL programme for extensive discussions and for sharing insights from ongoing STM/STS and x-ray work. D.A.R.\ and W.K.\ acknowledge support and facilities provided by the University of California, Berkeley.

\section*{Competing interests}
J.K.\ and H.K.\ are employees of Hyunsung T\&C\@. Hyunsung T\&C holds patent applications related to aspects of the MEL framework described in this work. D.A.R.\ and W.K.\ are affiliated with SuperMatics, Inc., which holds an exclusive licence from Hyunsung T\&C covering commercialization of the MEL framework. The authors declare no other competing interests.

\section*{Data availability}
The Fig.~\ref{fig:mc} data, lattice simulation code and plotting scripts are provided as ancillary files. The central estimates in Fig.~\ref{fig:mc} and Table~\ref{tab:mc_provenance} and the local-correlation values reproduce Table~II of the second version of Ref.~\cite{mel2025}; the condensate-gradient scan is an additional numerical result of the same model included in this study.

\clearpage
\appendix
\section*{Appendices}
\section{Variation of the Complex Fields}\label{app:variation}
The independent complex variations of Eq.~\eqref{F} are evaluated with vanishing surface terms. For the electronic envelope, $\partial|\Phi|^2/\partial\Phi^*=\Phi$ and $\partial|\nabla\Phi|^2/\partial(\partial_j\Phi^*)=\partial_j\Phi$. The gradient contribution is therefore $-\nabla\cdot[(K_m+\kappa|\psi|^2)\nabla\Phi]$. The condensate variation analogously produces $-D_j[(K_s+\kappa_\psi|\Phi|^2)D_j\psi]$. The pinning contribution $-\mathrm{Re}(h^*\Phi)$ gives $-h/2$. These operations yield the coupled stationary equations stated in the main text.

\section{Quartic and Gradient Stability}\label{app:stability}
Write $x=|\psi|^2\geq0$ and $y=|\Phi|^2\geq0$. The quartic part is
\begin{equation}\label{app:quartic}
Q_4(x,y)=\frac{b_s}{2}x^2+\gamma xy+\frac{b_m}{2}y^2.
\end{equation}
For $b_s,b_m>0$, $Q_4$ is strictly positive on the nonnegative quadrant away from the origin if $\gamma> -\sqrt{b_sb_m}$. At equality, the quartic form has a flat ray; boundedness of the full local potential along that ray depends on its quadratic coefficients. The interior coexistence Hessian is positive definite if $D=b_sb_m-\gamma^2>0$, with $x,y>0$ for the stationary solution. These conditions characterize different properties: quartic coercivity, local interior stability, and existence of the physical coexistence branch.

For the full continuum functional, the gradient contribution is Eq.~\eqref{grad_full}. Assume a bounded periodic domain, $K_s,K_m>0$, $h\in L^2(\Omega)$, a fixed regular gauge field and $b_s,b_m>0$ with $\gamma> -\sqrt{b_sb_m}$. There is then a constant $c>0$ such that
\begin{equation}\label{quartic_coercive}
Q_4(x,y)\geq c(x^2+y^2),\qquad x,y\geq0.
\end{equation}
The quadratic masses and pinning term can be bounded from below using Young's inequality, e.g., $|h||\Phi|\leq\varepsilon|\Phi|^4+C_\varepsilon|h|^{4/3}$, with $h\in L^2(\Omega)$ implying $h\in L^{4/3}(\Omega)$ on a finite domain. For $\kappa,\kappa_\psi\geq0$ the gradient part is nonnegative, so the complete functional is bounded below. This is a sufficient condition, not a classification of all stabilized extensions.

To see the obstruction for $\kappa<0$, choose $\psi=M$ spatially constant and $|M|^2>K_m/|\kappa|$ in any fixed regular gauge field, and let $\Phi=\varepsilon\sin(kx)$ with $k=2\pi n/L_x$. The negative gradient contribution scales as $(K_m+\kappa|M|^2)\varepsilon^2 k^2$, whereas the remaining terms stay bounded as $n\to\infty$ at fixed $M$ and $\varepsilon$. Hence $F\to-\infty$. For $\kappa_\psi<0$, take $\Phi=M$ constant with $|M|^2>K_s/|\kappa_\psi|$ and real $\psi=\varepsilon\sin(kx)$; then
\[
|\mathbf D\psi|^2=\varepsilon^2\big[k^2\cos^2kx+(q_s/\hbar)^2|\mathbf A|^2\sin^2kx\big],
\]
and $F\to-\infty$ as $n\to\infty$. Thus a positive gradient coefficient at a particular stationary point establishes only local stability; the unrestricted quadratic-gradient continuum functional with a negative mixed-gradient coefficient is not globally bounded below.

For the finite square lattice of Appendix~\ref{app:lattice}, put $x_i=|\psi_i|^2$, $y_i=|\Phi_i|^2$, $q=|\kappa_\psi|$ and assume $\kappa_\psi<0$, $\kappa\geq0$. The inequality $|\psi_i-\psi_j|^2\leq2(x_i+x_j)$ and the square-lattice adjacency norm $\|\mathcal A_{\rm nn}\|\leq4$ give
\[
E_4\geq \tfrac{b_s}{2}X^2+\tfrac{b_m}{2}Y^2
-[8q+\max(-\gamma,0)]XY,\qquad
X=\|x\|_2,\quad Y=\|y\|_2.
\]
Consequently $8|\kappa_\psi|+\max(-\gamma,0)<\sqrt{b_sb_m}$ ensures quartic coercivity and a finite lower bound at any fixed finite lattice and pinning realization. On an even periodic lattice, a staggered $\psi_i=R(-1)^{i_x+i_y}$ and constant $\Phi_i=\sqrt{u}R$ give
\[
E_4/N=R^4\big[\tfrac{b_s}{2}+(\gamma+8\kappa_\psi)u+\tfrac{b_m}{2}u^2\big],
\]
so $\gamma+8\kappa_\psi\geq-\sqrt{b_sb_m}$ is necessary. With the simulation parameters and $|\gamma|\leq0.3$, the sufficient threshold is $|\kappa_\psi|<(\sqrt{8}-0.3)/8\simeq0.316$; the scan value $-0.30$ satisfies it. This bound relies on the finite lattice cutoff and does not alter the continuum obstruction.

\section{Uniform-Limit Stiffness Response}\label{app:stiffness}
For the uniform local potential in Eq.~\eqref{uniform}, let $x_0=-a_s/b_s>0$ denote the squared condensate amplitude in the decoupled state at the same temperature. On an interior coexistence branch, the first stationarity condition gives
\begin{equation}\label{app:stationarity}
x=x_0-\frac{\gamma}{b_s}y,\qquad x=|\psi|^2,\quad y=|\Phi|^2.
\end{equation}
The uniform fixed-amplitude phase-gradient coefficient defines $\Upsilon_{\mathrm{bare}}=2(K_s+\kappa_\psi y)x$, with reference $\Upsilon_{0,\mathrm{bare}}=2K_sx_0$. Their ratio is
\begin{equation}\label{app:ratio}
R_{\mathrm{bare}}=\frac{x}{x_0}\left(1+\frac{\kappa_\psi y}{K_s}\right).
\end{equation}
Substitution of Eq.~\eqref{app:stationarity} yields the exact factorization
\begin{equation}\label{app:enhancement}
R_{\mathrm{bare}}-1=\frac{y}{K_sx_0}\left[\kappa_\psi x_0-\frac{K_s\gamma}{b_s}-\frac{\kappa_\psi\gamma}{b_s}y\right].
\end{equation}
For $K_s,x_0,x,y>0$, positive local gradient coefficients, and a stable interior branch, the sign of the bracket determines whether the homogeneous bare stiffness is enhanced. The $\kappa_\psi=0$ limit reduces to the condition $\gamma<0$, and the $\gamma=0$ limit to $\kappa_\psi>0$ when $y>0$. The coupling $\kappa$ does not enter this uniform expression because $\nabla\Phi=0$.

Equation~\eqref{app:enhancement} is an identity within the homogeneous fixed-amplitude reduction; the stiffness of a disordered, fluctuating system follows from Eq.~\eqref{app:twist_eq}.

\section{Helicity Modulus from Boundary Twists}\label{app:twist}
For a total phase twist $\delta$ across a rectangular two-dimensional domain, let $\mathcal F(\delta)=-k_BT\log Z(\delta)$ as in Sec.~\ref{sec:helicity}. The helicity modulus is
\begin{equation}\label{app:twist_eq}
\Upsilon_x=\frac{L_x}{L_y}\left.\frac{\partial^2\mathcal F}{\partial\delta^2}\right|_{\delta=0}
=\frac{L_x}{L_y}\left[\left\langle F_\delta''\right\rangle-\beta\left(\left\langle(F_\delta')^2\right\rangle-\left\langle F_\delta'\right\rangle^2\right)\right]_{\delta=0}.
\end{equation}
The derivatives are taken with the complete twist-dependent lattice bond energy. In particular, the $\kappa_\psi|\Phi|^2$ contribution to the bond weight enters both the energy curvature and current-fluctuation terms. The fixed-amplitude result in Appendix~\ref{app:stiffness} is recovered when these fluctuations are suppressed.

\section{Lattice Monte Carlo Model}\label{app:lattice}
The numerical model uses complex site fields on an $L=120$ periodic square lattice, with the lattice spacing as unit length and $\mathbf A=0$. With $x_i=|\psi_i|^2$, $y_i=|\Phi_i|^2$ and $f_0$ from Eq.~\eqref{uniform}, its energy is
\begin{align*}
E={}&\sum_i\big[f_0(x_i,y_i)-\Ree(h_i^*\Phi_i)\big]\\
&+\sum_{\langle ij\rangle}\Big\{
\big[K_s+\tfrac{\kappa_\psi}{2}(y_i+y_j)\big]|\psi_i-\psi_j|^2\\
&\hspace{4.5em}+
\big[K_m+\tfrac{\kappa}{2}(x_i+x_j)\big]|\Phi_i-\Phi_j|^2
\Big\}.
\end{align*}
Each nearest-neighbour bond is counted once; both mixed-gradient weights use the symmetric average of the squared field at the bond endpoints. The parameters are $a_s(T)=a_{s0}(T/T^*-1)$, $a_{s0}=4$, $T^*=2.5$, $b_s=4$, $K_s=1$, $a_m=-1$, $b_m=2$ and $K_m=2$, in units of $\varepsilon_0$ with $k_B=1$. The quenched pinning variables are independent circular complex Gaussians, $h_i=(W/\sqrt2)(\eta_i+i\zeta_i)$, with independent standard real normal variables $\eta_i,\zeta_i$ and $W=0.6$; thus $\mathbb E|h_i|^2=W^2$, and each Cartesian component has standard deviation $W/\sqrt2$. The reference model sets $\gamma=\kappa=\kappa_\psi=0$ at the same $T,W$ and disorder statistics, and the crossing of $\Upsilon(T)$ at $L=120$ with $2T/\pi$ defines $T_{c0}=0.785$; the scan in Table~\ref{tab:kpsi} instead keeps $\kappa_\psi$ equal in numerator and reference.

Each sweep visits all sites sequentially, updating $\psi_i$ and then $\Phi_i$ by complex Metropolis proposals followed by a microcanonical phase reflection about the corresponding local field. This over-relaxation is applied to both fields and preserves the local energy. The cooling grid runs from $T=1.30$ to $0.30$; each temperature starts from the final configuration at the previous one, after 100 step-size tuning sweeps without over-relaxation, and uses $4000$ equilibration sweeps and $12000$ measurement sweeps, recording every sixth sweep. $\Upsilon$ is Eq.~\eqref{app:twist_eq} evaluated with the lattice bond energy and averaged over both directions; the Pearson coefficient is computed over sites in each recorded configuration and then averaged. The chain counts for Fig.~\ref{fig:mc} are in Table~\ref{tab:mc_provenance}; the nonzero-$\kappa_\psi$ scan uses three independent chains per point. The $1.9\%$ history systematic comes from a comparison of three heating chains with the six cooling chains at $(\gamma,\kappa)=(-0.3,1)$, while complementary finite-size checks give reference crossings $0.789$ at $L=60$ and $0.784$ at $L=160$, with stiffness ratios at $(\gamma,\kappa)=(\mp0.3,1)$ remaining $1.00$ and $0.87$ at both sizes.


\clearpage
\begin{thebibliography}{99}
\bibitem{mel2025} J. Kim, D. A. Rens, W. Khalid, and H. Kim, ``Short-Range Modulated Electron Lattice and d-Wave Superconductivity in Cuprates: A Phenomenological Ginzburg--Landau Framework,'' arXiv:2512.03368 (v1, 2025; v2, 2026).
\bibitem{timusk1999} T. Timusk and B. Statt, ``The pseudogap in high-temperature superconductors: An experimental survey,'' Rep. Prog. Phys. 62, 61 (1999).
\bibitem{keimer2015} B. Keimer et al., ``From quantum matter to high-temperature superconductivity in copper oxides,'' Nature 518, 179 (2015).
\bibitem{fradkin2015} E. Fradkin, S. A. Kivelson, and J. M. Tranquada, ``Colloquium: Theory of intertwined orders in high temperature superconductors,'' Rev. Mod. Phys. 87, 457 (2015).
\bibitem{damascelli2003} A. Damascelli, Z. Hussain, and Z.-X. Shen, ``Angle-resolved photoemission studies of the cuprate superconductors,'' Rev. Mod. Phys. 75, 473 (2003).
\bibitem{norman1998} M. R. Norman et al., ``Destruction of the Fermi surface in underdoped high-$T_c$ superconductors,'' Nature 392, 157 (1998).
\bibitem{fischer2007} \O. Fischer et al., ``Scanning tunneling spectroscopy of high-temperature superconductors,'' Rev. Mod. Phys. 79, 353 (2007).
\bibitem{ghiringhelli2012} G. Ghiringhelli et al., ``Long-range incommensurate charge fluctuations in (Y,Nd)Ba$_2$Cu$_3$O$_{6+x}$,'' Science 337, 821 (2012).
\bibitem{chang2012} J. Chang et al., ``Direct observation of competition between superconductivity and charge density wave order in YBa$_2$Cu$_3$O$_{6.67}$,'' Nat. Phys. 8, 871 (2012).
\bibitem{comin2016} R. Comin and A. Damascelli, ``Resonant x-ray scattering studies of charge order in cuprates,'' Annu. Rev. Condens. Matter Phys. 7, 369 (2016).
\bibitem{hoffman2002} J. E. Hoffman et al., ``A four unit cell periodic pattern of quasi-particle states surrounding vortex cores in Bi$_2$Sr$_2$CaCu$_2$O$_{8+\delta}$,'' Science 295, 466 (2002).
\bibitem{vershinin2004} M. Vershinin et al., ``Local ordering in the pseudogap state of the high-$T_c$ superconductor Bi$_2$Sr$_2$CaCu$_2$O$_{8+\delta}$,'' Science 303, 1995 (2004).
\bibitem{wise2008} W. D. Wise et al., ``Charge-density-wave origin of cuprate checkerboard visualized by scanning tunnelling microscopy,'' Nat. Phys. 4, 696 (2008).
\bibitem{reznik2006} D. Reznik et al., ``Electron-phonon coupling reflecting dynamic charge inhomogeneity in copper oxide superconductors,'' Nature 440, 1170 (2006).
\bibitem{letacon2014} M. Le Tacon et al., ``Inelastic x-ray scattering in YBa$_2$Cu$_3$O$_{6.6}$ reveals giant phonon anomalies and elastic central peak due to charge-density-wave formation,'' Nat. Phys. 10, 52 (2014).
\bibitem{sonier1994} J. E. Sonier et al., ``New muon-spin-rotation measurement of the temperature dependence of the magnetic penetration depth in YBa$_2$Cu$_3$O$_{6.95}$,'' Phys. Rev. Lett. 72, 744 (1994).
\bibitem{sonier2007} J. E. Sonier et al., ``Hole-doping dependence of the magnetic penetration depth and vortex core size in YBa$_2$Cu$_3$O$_y$: Evidence for stripe correlations near 1/8 hole doping,'' Phys. Rev. B 76, 134518 (2007).
\bibitem{pumpin1990} B. P\"umpin et al., ``Muon-spin-rotation measurements of the London penetration depths in YBa$_2$Cu$_3$O$_{6.97}$,'' Phys. Rev. B 42, 8019 (1990).
\bibitem{blatter1994} G. Blatter et al., ``Vortices in high-temperature superconductors,'' Rev. Mod. Phys. 66, 1125 (1994).
\bibitem{agterberg2020} D. F. Agterberg et al., ``The physics of pair-density waves: Cuprate superconductors and beyond,'' Annu. Rev. Condens. Matter Phys. 11, 231 (2020).
\bibitem{fujita2014} K. Fujita et al., ``Direct phase-sensitive identification of a d-form factor density wave in underdoped cuprates,'' Proc. Natl. Acad. Sci. USA 111, E3026 (2014).
\bibitem{hayward2014} L. E. Hayward et al., ``Angular fluctuations of a multicomponent order describe the pseudogap of YBa$_2$Cu$_3$O$_{6+x}$,'' Science 343, 1336 (2014).
\bibitem{caplan2015} Y. Caplan, G. Wachtel, and D. Orgad, ``Long-range order and pinning of charge-density waves in competition with superconductivity,'' Phys. Rev. B 92, 224504 (2015).
\bibitem{yukivelson2019} Y. Yu and S. A. Kivelson, ``Fragile superconductivity in the presence of weakly disordered charge density waves,'' Phys. Rev. B 99, 144513 (2019).
\bibitem{morozovska2023} A. N. Morozovska et al., ``Landau--Ginzburg theory of charge density wave formation accompanying lattice and electronic long-range ordering,'' Phys. Rev. B 107, 174104 (2023).
\bibitem{nie2015} L. Nie et al., ``Fluctuating orders and quenched randomness in the cuprates,'' Phys. Rev. B 92, 174505 (2015).
\bibitem{kivelson2002} S. A. Kivelson et al., ``Competing order in the mixed state of high-temperature superconductors,'' Phys. Rev. B 66, 144516 (2002).
\bibitem{mesaros2016} A. Mesaros et al., ``Commensurate $4a_0$-period charge density modulations throughout the Bi$_2$Sr$_2$CaCu$_2$O$_{8+x}$ pseudogap regime,'' Proc. Natl. Acad. Sci. USA 113, 12661 (2016).
\bibitem{slezak2008} J. A. Slezak et al., ``Imaging the impact on cuprate superconductivity of varying the interatomic distances within individual crystal unit cells,'' Proc. Natl. Acad. Sci. USA 105, 3203 (2008).
\bibitem{orgad2024} D. Orgad, ``Disorder effects in a model of competing superconducting and charge-density wave orders in YBa$_2$Cu$_3$O$_{6+x}$,'' arXiv:2409.16043 (2024).
\bibitem{caplan2017} Y. Caplan and D. Orgad, ``Dimensional crossover of charge-density wave correlations in the cuprates,'' Phys. Rev. Lett. 119, 107002 (2017).
\bibitem{delmaestro2006} A. Del Maestro et al., ``From stripe to checkerboard ordering of charge-density waves on the square lattice in the presence of quenched disorder,'' Phys. Rev. B 74, 024520 (2006).
\bibitem{robertson2006} J. A. Robertson et al., ``Distinguishing patterns of charge order: Stripes or checkerboards,'' Phys. Rev. B 74, 134507 (2006).
\bibitem{fisher1973} M. E. Fisher, M. N. Barber, and D. Jasnow, ``Helicity modulus, superfluidity, and scaling in isotropic systems,'' Phys. Rev. A 8, 1111 (1973).
\bibitem{liufisher1973} K.-S. Liu and M. E. Fisher, ``Quantum lattice gas and the existence of a supersolid,'' J. Low Temp. Phys. 10, 655 (1973).
\bibitem{knf1976} J. M. Kosterlitz, D. R. Nelson, and M. E. Fisher, ``Bicritical and tetracritical points in anisotropic antiferromagnetic systems,'' Phys. Rev. B 13, 412 (1976).
\bibitem{imryma1975} Y. Imry and S.-k. Ma, ``Random-field instability of the ordered state of continuous symmetry,'' Phys. Rev. Lett. 35, 1399 (1975).
\end{thebibliography}
\end{document}